\pageno=1 
\baselineskip=20pt 
\magnification=\magstep1
\def\bs{\bigskip}
\def\ni{\noindent} 
\def\ms{\medskip} 
\def\eq{\eqno}
\input epsf

\bs\bs 

\centerline {\bf NUCLEAR MATTER IN RELATIVISTIC MEAN FIELD THEORY} 

\centerline {\bf WITH ISOVECTOR SCALAR MESON} 

\bs\bs \bs\bs

\centerline {S. Kubis and M. Kutschera} 

\centerline {H.Niewodnicza\'nski Institute of Nuclear Physics} 

\centerline {ul. Radzikowskiego 152, 31-342 Krak\'ow, Poland} 

\bs \bs \bs \bs 

\ni Abstract:

Relativistic mean field (RMF) theory of nuclear matter with the isovector
scalar mean field corresponding to the $\delta$-meson [$a_0(980)$] is studied. 
While the $\delta$-meson mean field vanishes in symmetric nuclear matter, it
can influence properties of asymmetric nuclear matter
in neutron stars. 
The RMF contribution due to $\delta$-field to the nuclear
symmetry energy is negative. To fit the empirical value,
$E_s \approx 30 MeV$, 
a stronger $\rho$-meson coupling is required than in the absence of
the $\delta$-field.
The energy per particle of
neutron matter is then larger at high densities than the one
with no $\delta$-field included. Also, the proton fraction of $\beta$-stable
matter increases. Splitting of proton and neutron effective
masses due to the $\delta$-field can affect transport properties
of neutron star matter.
\ms

\ni PACS: 21.65.+f, 97.60.Jd

\ni {\it key words: nuclear matter, relativistic mean field theory, isovector 
scalar exchange, nuclear symmetry energy, neutron star equation of state,
nucleon effective mass}

\bs \bs

RMF models of nuclear matter [1] play an
important role in dense matter calculations, where relativistic
effects increase with density.  They represent a relativistic
Hartree approximation to the one-boson exchange (OBE) theory of nuclear
interactions, with appropriately readjusted couplings to fit the
saturation properties of nuclear matter. 
In the OBE description of nucleon-nucleon scattering,
isoscalar mesons,
$\sigma$, $\eta$, and $\omega$, and isovector
mesons, $\pi$, $\rho$, and $\delta$, are involved.

In the RMF approximation, mean fields of pseudoscalar
mesons $\eta$ and $\pi$ vanish  in normal nuclear 
matter.
The standard RMF model of nuclear matter [1] 
includes isoscalar mesons $\sigma$ and  
$\omega$, and   the isovector $\rho$-meson. However,
no $\delta$-meson contribution is present, although
for asymmetric nuclear matter the density to which 
the $\delta$-field can couple does not vanish,
$<{\bar\psi}\tau_3\psi> \ne 0$. 

The $\delta$-field is not expected to play an important role for
nuclei, whose isospin asymmetry is not large. Also, the short
range of the $\delta$-meson exchange justifies neglecting its contribution at
saturation density. However, for strongly isospin-asymmetric
matter at high densities in neutron stars the contribution of the
$\delta$-field should be considered.
It is our aim to include the $\delta$-meson into 
the RMF model and study consequences of this generalization of the RMF
theory.  To our knowledge, such an extension of RMF models has
not been studied.

We assume here that all the meson fields have the Yukawa couplings to
nucleons.  The interaction lagrangian reads

$$ L_{int}=g_{\sigma} \sigma {\bar \psi} \psi - g_{\omega}
\omega_{\mu} {\bar \psi} \gamma^{\mu} \psi -{1 \over 2}g_{\rho} {\vec \rho}_{\mu}
{\bar \psi} \gamma^{\mu} {\vec \tau} \psi +g_{\delta} {\vec \delta}
{\bar \psi}{\vec \tau} \psi.  \eq(1)$$

\ni Here ${\vec \delta}$ is the isovector scalar field of the
$\delta$-meson.  The free field lagrangians for $\psi, \sigma, \omega$,
and $\rho$ fields are the same as in Ref.[1].  For the $\delta$-field
we use the standard lagrangian

$$L_{\delta}={1 \over 2} \partial_{\mu}{\vec
\delta}\partial^{\mu}{\vec \delta}-{1 \over 2}m_{\delta}^2{\vec
\delta}^2.  \eq(2)$$

\ni For the $\sigma$-field we adopt the potential energy term of Boguta
and Bodmer [2],

$$ U(\sigma)= {1 \over 3} b m\sigma^3 + {1 \over 4} c \sigma^4,
\eq(3)$$

\ni where $m$ is the bare nucleon mass.

The relevant components of meson fields are $\sigma$, $\omega_0$, 
$\rho_0^{(3)}$, and
for the $\delta$-field the isospin component $\delta^{(3)}$.
All remaining components vanish, in
particular $\delta^{(1)}=\delta^{(2)}=0$. 
The values
of mean fields in the ground state are
determined by proton and neutron densities.  
The field equations for vector meson fields give ${\bar
\omega}_0=(g_{\omega}/m_{\omega}^2)n_B$ and

\ni ${\bar \rho}_0^{(3)}=(g_{\rho}/m_{\rho}^2)(2x-1)n_B$, where
$n_B$ is the baryon density and
$x=n_P/n_B$ is the proton fraction.  The
field equations for scalar fields are non-trivial:

$$m_{\sigma}^2{\bar \sigma}+{\partial U \over \partial \sigma}=g_{\sigma}(n^s_P+n^s_N), \eq(4)$$

\ni and

$$m_{\delta}^2{\bar \delta^{(3)}}=g_{\delta}(n^s_P-n^s_N).  \eq(5)$$

\ni Here $n^s_P$ and $n^s_N$ is, respectively, proton and neutron
scalar density:

$$ n^s_i={2 \over (2\pi)^3} \int^{k_i}_0d^3k {m_i \over
\sqrt{k^2+m_i^2}}, ~~~~~~~~i=P,N.\eq(6)$$

\ni In Eq.(6)
$m_P$ and $m_N$ is, respectively, proton
and neutron effective mass,

$$ m_P=m-g_{\sigma}{\bar \sigma}-g_{\delta}{\bar \delta^{(3)}}, \eq(7) $$

\ni and

$$ m_N=m-g_{\sigma}{\bar \sigma}+g_{\delta}{\bar \delta^{(3)}}. \eq(8) $$

The energy density of uniform nucleon matter is

$$ \epsilon = {2 \over (2\pi)^3} (\int_0^{k_P}d^3k
\sqrt{k^2+m_P^2}+\int_0^{k_N}d^3k \sqrt{k^2+m_N^2})+ {1 \over
2}C_{\omega}^2n_B^2 + $$

$$ +{1 \over 2} {1 \over C_{\sigma}^2}[m-{m_P+m_N \over 2}]^2+ U({\bar
\sigma})+{1 \over 8}C_{\rho}^2(2x-1)^2n_B^2+{1 \over 8} {1 \over
C_{\delta}^2}(m_N-m_P)^2.  \eq(9)$$

The model parameters in the isoscalar sector,
$C_{\sigma}^2 \equiv g_{\sigma}^2/m_{\sigma}^2,
C_{\omega}^2 \equiv g_{\omega}^2/m_{\omega}^2, {\bar b} \equiv
b/g_{\sigma}^3$, and ${\bar c} \equiv c/g_{\sigma}^3$, are adjusted to 
fit the saturation properties of symmetric nuclear matter, the
saturation density $n_0=0.145fm^{-3}$, the binding energy $w_0=-16
MeV$ per nucleon, and the compressibility modulus $K_V \approx 280MeV$.
The fourth parameter, e.g. ${\bar c}$, can be used to measure stiffness of
the equation of state of symmetric nuclear matter.  In the following
we use two sets of parameters which reproduce the saturation
properties but differ at higher densities.  The soft
equation of state is specified by the parameters
$C_{\sigma}^2=1.582fm^2, C_{\omega}^2=1.019fm^2, {\bar
b}=-0.7188$, and ${\bar c}=6.563$.
For the stiff equation of state  the parameters are
$C_{\sigma}^2=11.25fm^2, C_{\omega}^2=6.483fm^2, {\bar b}=0.003825$, and
${\bar c}=3.5\times10^{-6}$. 

In the spirit of RMF models,
the parameters $C_{\rho}^2 \equiv g_{\rho}^2/m_{\rho}^2 $ and
$C_{\delta}^2 \equiv g_{\delta}^2/m_{\delta}^2$ of the isovector 
sector should be constrained to fit the 
nuclear symmetry energy, $E_s=31\pm 4 MeV$ [3].  In terms of the model
parameters, the symmetry energy is

$$ E_s={1 \over 8}C_{\rho}^2n_0+{k_0^2 \over 6\sqrt{k_0^2+m_0^2}}
-C_{\delta}^2 {m_0^2 n_0 \over 2(k_0^2+m_0^2)(1+C_{\delta}^2A(k_0,m_0))},
 \eq(10) $$

\ni where

$$ A(k_0,m_0)={4\over (2\pi)^3} \int_0^{k_0} {d^3p p^2 \over (p^2+m_0^2)^{3/2}}
\eq(11)$$

\ni is a function of the Fermi momentum,
$k_0=k_P=k_N$, and the effective mass, $m_0=m_P=m_N$, of symmetric nuclear
matter at saturation density.
The constraint (10) gives 
$C_{\rho}^2$ as a function of $C_{\delta}^2$,
$C_{\rho}^2\equiv C_{\rho}^2(C_{\delta}^2)$, which is shown in Fig.1. Hence by
demanding to fit the bulk properties of nuclear matter we are not able
to fix independently $C_{\rho}^2$ and $C_{\delta}^2$. Instead, we
explore here a range of values of $C_{\delta}^2$ corresponding to the
Bonn potentials A, B and C [4]. The maximum
value $C_{\delta}^2=2.6fm^2$ corresponds to the parameter
$g_{\delta}=8.0$ of 
the Bonn potential C [4]. 

The fact that the constraint (10) gives $C_{\rho}^2$ as a
monotonically increasing function of $C_{\delta}^2$ plays a
crucial role in our analysis. The nuclear symmetry energy,
Eq.(10), is derived in the RMF approximation which neglects
exchange contributions. Inclusion of the Fock terms in 
relativistic nuclear matter calculations gives a sizable
correction to the nuclear symmetry energy [5]. We expect,
however, that adding the Fock contribution will not change the
general behaviour of the curves in Fig.1, although it can affect
their intercept and slope. This topic will be discussed in
detail elsewhere [6].

To estimate the effect of the $\delta$-field we first consider
the case when the coupling constant $C_{\rho}^2$ is adjusted
with no $\delta$-field present ($C_{\delta}^2=0$). For the soft
and stiff equation of state we find, respectively,
$C_{\rho}^2=5.0fm^2$ and $C_{\rho}^2=4.29fm^2$. In Fig.2 we show
corresponding energies of pure neutron matter. When the 
$\delta$-field is switched on, the energy/particle becomes
smaller. In Fig.2 we show curves for a few values of
$C_{\delta}^2$ from the range compatible with the 
Bonn potentials A, B and C [4]. 

The contribution due to the
$\delta$-field is attractive and quite large. In case of the
stiff equation of state the neutron matter becomes selfbound for 
$C_{\delta}^2 \ge 1.1fm^2$.  
Already a smaller value, $C_{\delta}^2 \approx 1.0fm^2$, lowers the
symmetry energy to about $20 MeV$. One should note that for
$C_{\rho}^2=2.5fm^2$ the neutron matter becomes as strongly
bound as the symmetric nuclear matter. 
For the soft equation of state the binding is stronger.

To avoid such an unphysical situation, the
increased binding due to the $\delta$-field has
to be balanced by the higher repulsion due to the $\rho$-field. In
particular, if we demand the symmetry energy to be reproduced,
the formula (10) shows that the parameter
$C_{\rho}^2$ has its lowest value for $C_{\delta}^2=0$. 
For any finite value of the $\delta$-coupling, $C_{\delta}^2> 0$, the
strength of the $\rho$-coupling, $C_{\rho}^2$, increases. In this
case inclusion of the 
$\delta$-meson results 
in higher energy/particle at high
densities, where the contribution of vector mesons dominates.
 The
value  $C_{\rho}^2(0)$ should be regarded as the lower bound, with the
actual value larger, corresponding to some finite $C_{\delta}^2$. Similarly, 
the energy/particle of neutron matter is bounded from below by values obtained
for $C_{\rho}^2(0)$. Actual energies, for finite $C_{\delta}^2$, are higher.
Below we show results for $C_{\delta}^2=2.5fm^2$ which is close
to the value of the ratio $g_{\delta}^2/m_{\delta}^2$ corresponding
to the Bonn potential C [4].

The presence of the mean $\delta$-field leads to splitting of
proton and neutron effective masses, Eqs.(7) and (8).
In Fig.3 and Fig.4
we show the effective masses $m_P$ and $m_N$ as functions of baryon
density for a few values of the proton fraction $x$.  It is
interesting to note that for the stiff equation of state, Fig.3,
the proton and 
neutron effective masses in pure neutron matter both decrease
monotonically with density.  For the soft equation of state,
Fig.4, the 
proton effective mass in pure neutron matter increases with density,
approaching asymptotically the value $m_P=2m$ for the largest
couplings $C_{\delta}^2$. In contrast, the neutron effective
mass decreases monotonically. Such a splitting of proton and
neutron mass can affect the transport properties of dense matter.

The energy per particle is presented in Fig.5 where we show
results for pure neutron matter      
for both soft and stiff equations of state.
For comparison, curves for neutron matter with no $\delta$-field
included, and for symmetric nuclear nuclear matter, are also
shown. One can notice that the energy/particle with the
$\delta$-contribution included increases more rapidly with
neutron matter 
density for both equations of state than the energy for
$C_{\delta}^2=0$.   

The proton fraction of $\beta$-stable neutron star matter, which
satisfies the condition

$$\mu_N-\mu_P = \sqrt{k_N^2+m_N^2}-\sqrt{k_P^2+m_P^2}+{1 \over
2}C_{\rho}^2(1-2x)n_B=\mu_e, \eq(12) $$ 

\ni where the electron chemical potential is
$\mu_e=(3\pi^2xn_B)^{1/3}$, is shown 
in Fig.6.  One can notice that for 
$C_{\delta}^2=2.5fm^2$ the proton fraction at high densities is
larger than for $C_{\delta}^2=0$ for both equations of state.
It exceeds the critical value for the direct URCA process to be of
importance in the cooling of neutron stars, which is 
$x_{URCA}\approx 0.11$.

In conclusion, the RMF theory with the contribution due to the
$\delta$-meson mean field, constrained to fit bulk properties of
nuclear matter, predicts higher energy/particle of the
neutron matter at a given baryon density than in absence of the
$\delta$-field. Although  
the $\delta$-field provides an additional binding, fitting of
the symmetry energy, $E_s \approx 30 MeV$, requires a stronger
$\rho$-meson coupling which dominates at higher densities. The
proton fraction of 
the $\beta$-stable neutron star matter is considerably higher
than the one with
no $\delta$-field contribution. 
Proton and neutron effective masses are
split, an effect which can modify the transport properties of
dense matter.

\bs

This research is partially supported by the Polish State
Committee for Scientific Research (KBN), grants 2 P03B 083 08
and 2 P03D 001 09.

\bs

\ni {\bf References}

\ni~~[1]~B. D. Serot and J. D. Walecka, Adv.  Nucl.
Phys. {\bf16} (1986) 1.

\ni~~[2]~J. Boguta and A. Bodmer, Nucl. Phys. {\bf A292} (1977) 413.

\ni~~[3]~W. D. Myers and W. D. Swiatecki, Ann. Phys. {\bf 84}
(1973) 186; 

\ni~~~~~~J. M. Pearson, Y. Aboussir, A. K. Dutta, R. C. Nayak,
M. Farine, and F. Tondeur, 

\ni~~~~~~Nucl. Phys. {\bf A528} (1991) 1;

\ni~~~~~~P. M\"oller and J. R. Nix, At. Data and Nucl. Data
Tables {\bf 39} (1988) 219;

\ni~~~~~~P. M\"oller, W. D. Myers, W. J. Swiatecki, and J.
Treiner, At. Data and Nucl. Data 

\ni~~~~~~Tables {\bf 39} (1988) 225;

\ni~~~~~~W. D. Myers, W. J. Swiatecki, T. Kodama, L. J.
El-Jaick, and E. R. Hilf, 

\ni~~~~~Phys. Rev. {\bf C15} (1977) 2032.

\ni~~[4]~R. Machleidt, Adv. Nucl. Phys. {\bf 19} (1989) 189.

\ni~~[5]~A. Bouyssy, J.-F. Mathiot, Nguyen Van Giai, and S.
Marcos, Phys. Rev. C {\bf 36},

\ni~~~~~~(1987) 380;

\ni~~~~~~M. Lopez-Quelle, S. Marcos, R. Niembro, A. Bouyssy, and
Nguyen Van Giai, Nucl. 

\ni~~~~~~Phys. {\bf A483} (1988) 479.

\ni~~[6]~S. Kubis and M. Kutschera, in preparation.  
\bs
\bs
%\ni {\bf Figure captions}

\epsffile {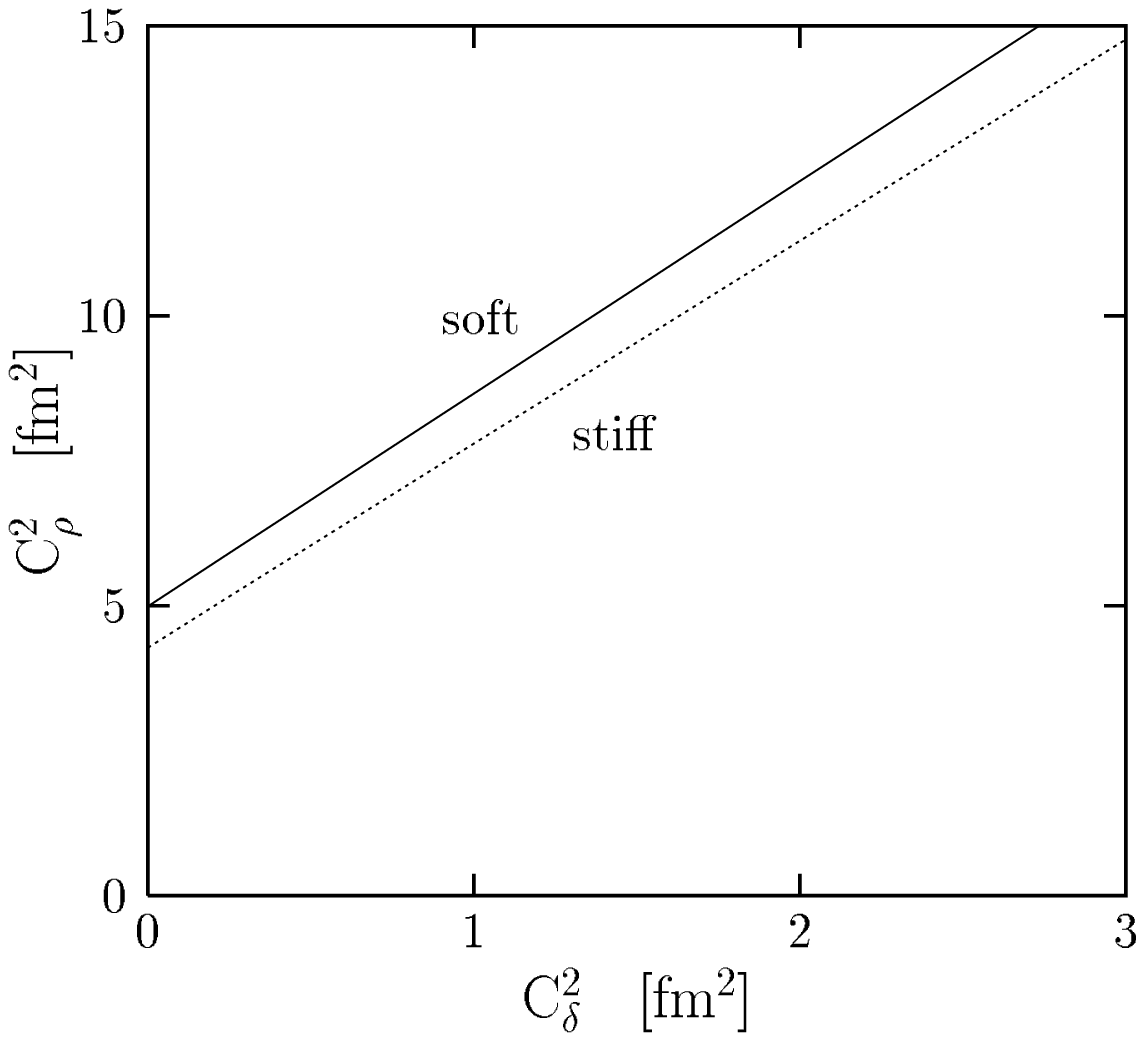}
\ni Fig.1

\ni The $\rho$-meson coupling, $C_{\rho}^2$, required to fit the
empirical value of the nuclear symmetry energy, $E_s\approx
30 MeV$, as a function of the $\delta$-meson coupling,
$C_{\delta}^2$. 

\vfill
\break

\epsffile {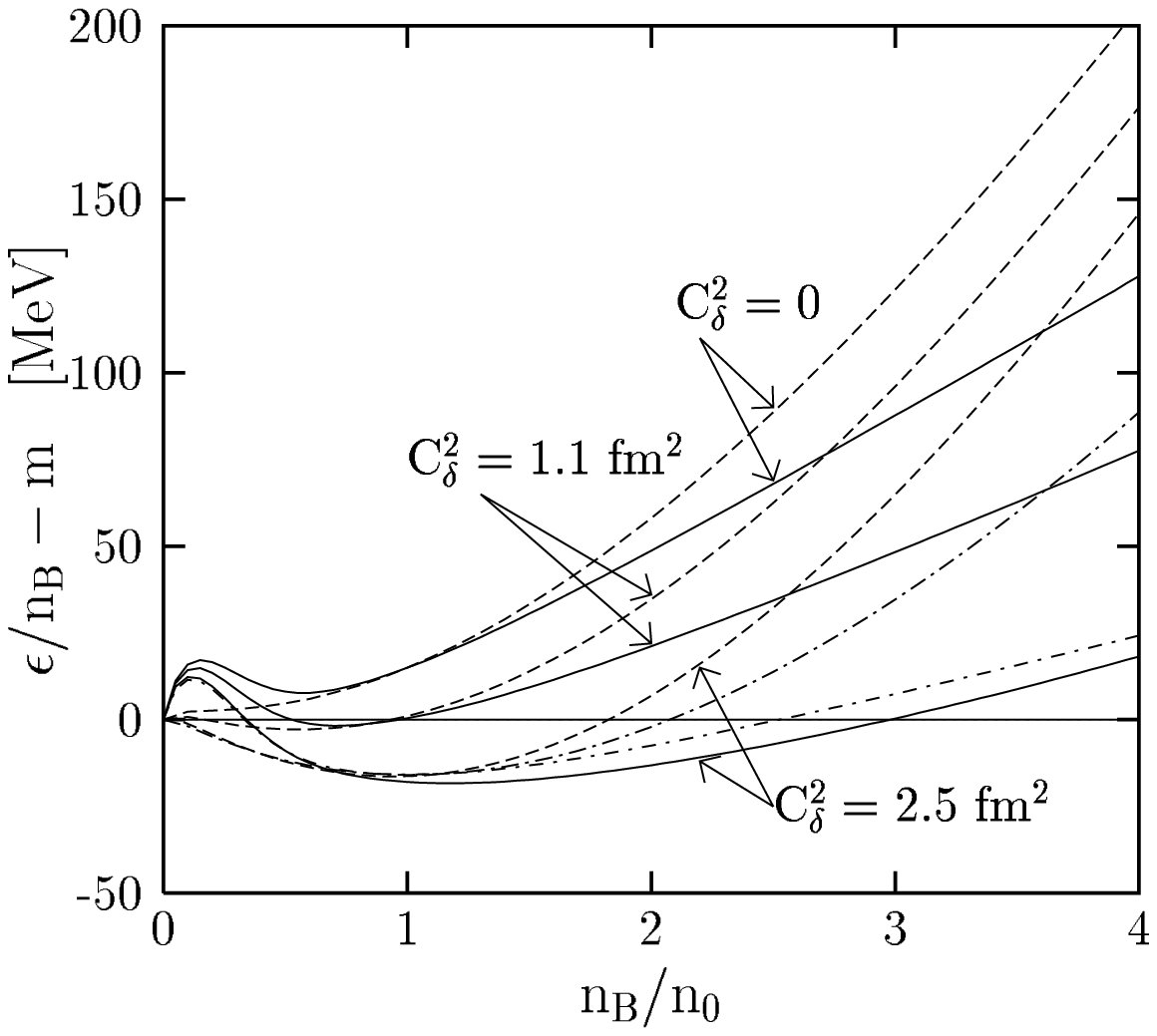}
\ni Fig.2

\ni The energy per particle of pure neutron matter for a few values of the
$\delta$-meson coupling, $C_{\delta}^2$. For
$C_{\rho}^2$ the same value is used for all curves; it is the
value which fits the nuclear symmetry energy with no
$\delta$-field. Solid and dashed curves correspond to the soft
and stiff equation of state, respectively. For comparison, the
energy per particle 
of symmetric nuclear matter is shown (dash-dotted curves).

\vfill
\break

\epsffile {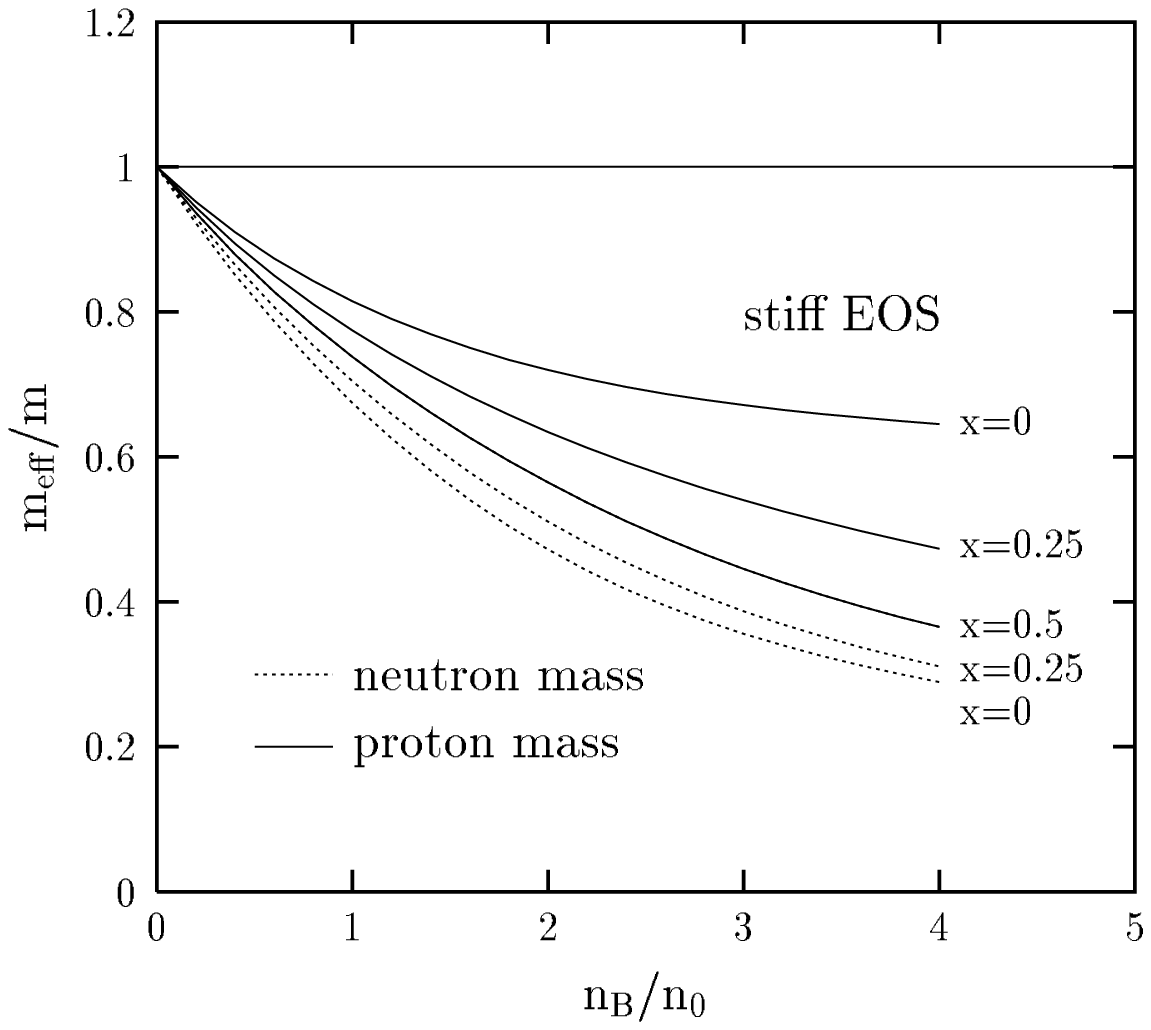}
\ni Fig.3

\ni Effective masses of protons  and neutrons 
for a few values of the proton fraction, for  the stiff
equation of state and for $C_{\delta}^2=2.5 fm^2$.

\vfill
\break

\epsffile {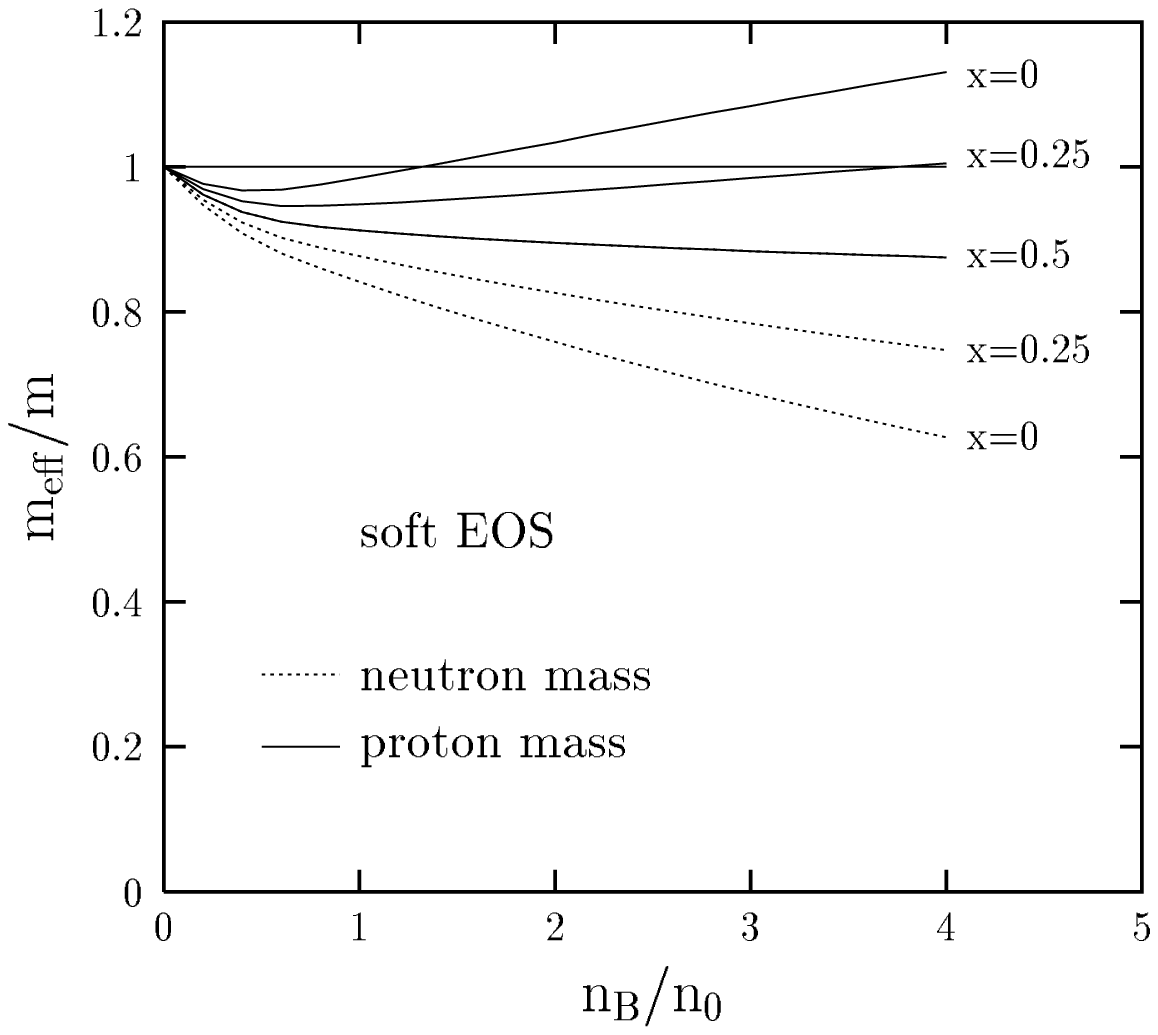}
\ni Fig.4

\ni The same as Fig.3 for the soft equation of state.

\vfill
\break

\epsffile {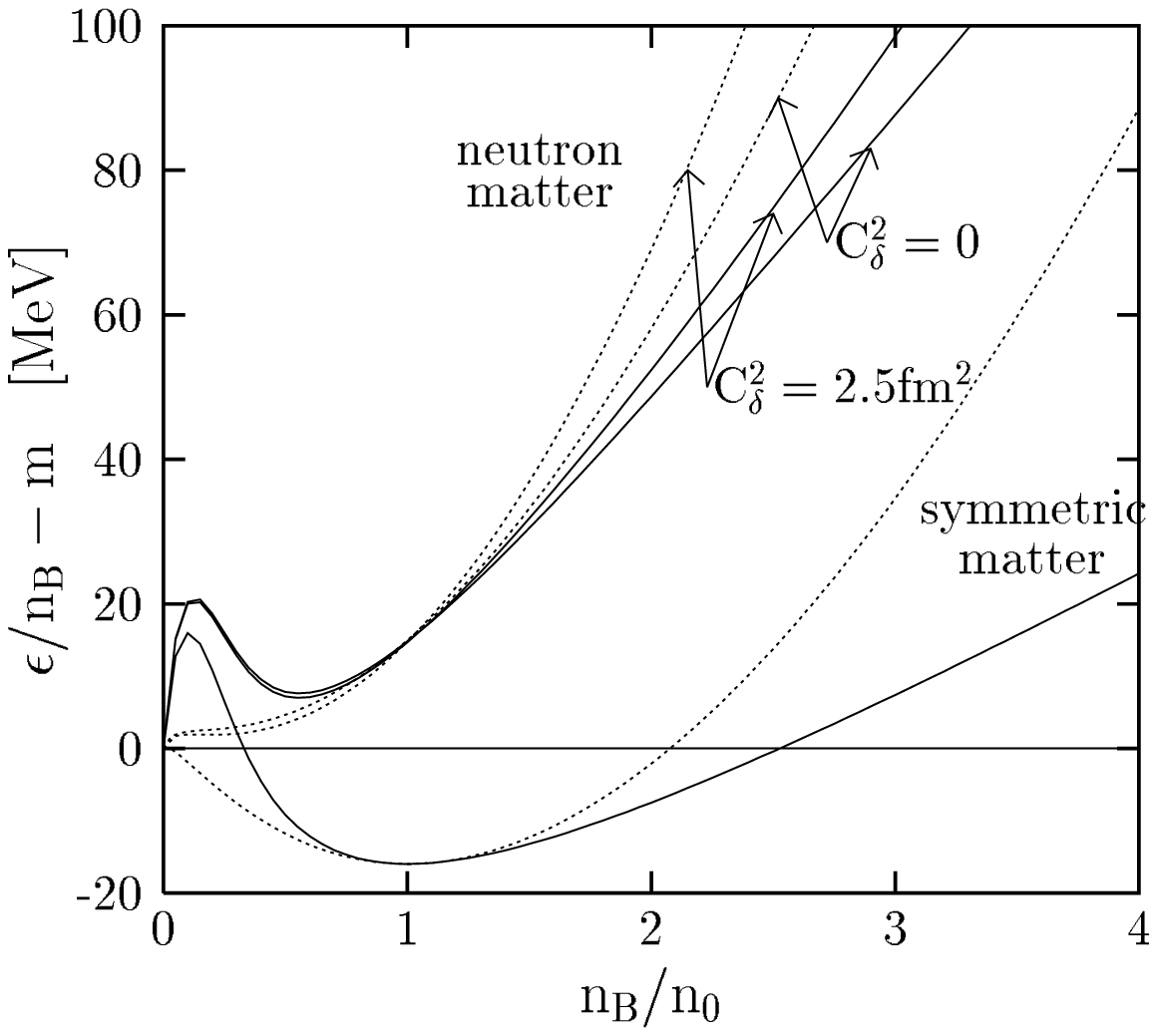}
\ni Fig.5

\ni The energy per particle of neutron matter for
$C_{\delta}^2=2.5 fm^2$, corresponding to 
the Bonn  potential C, and with no
$\delta$-meson contribution. The parameter $C_{\rho}^2$ is adjusted to
fit the nuclear symmetry energy in both cases. Solid and dotted
curves are, respectively, for the soft and stiff equation of
state. The energy per particle of symmetric nuclear matter is also shown.

\vfill
\break

\epsffile {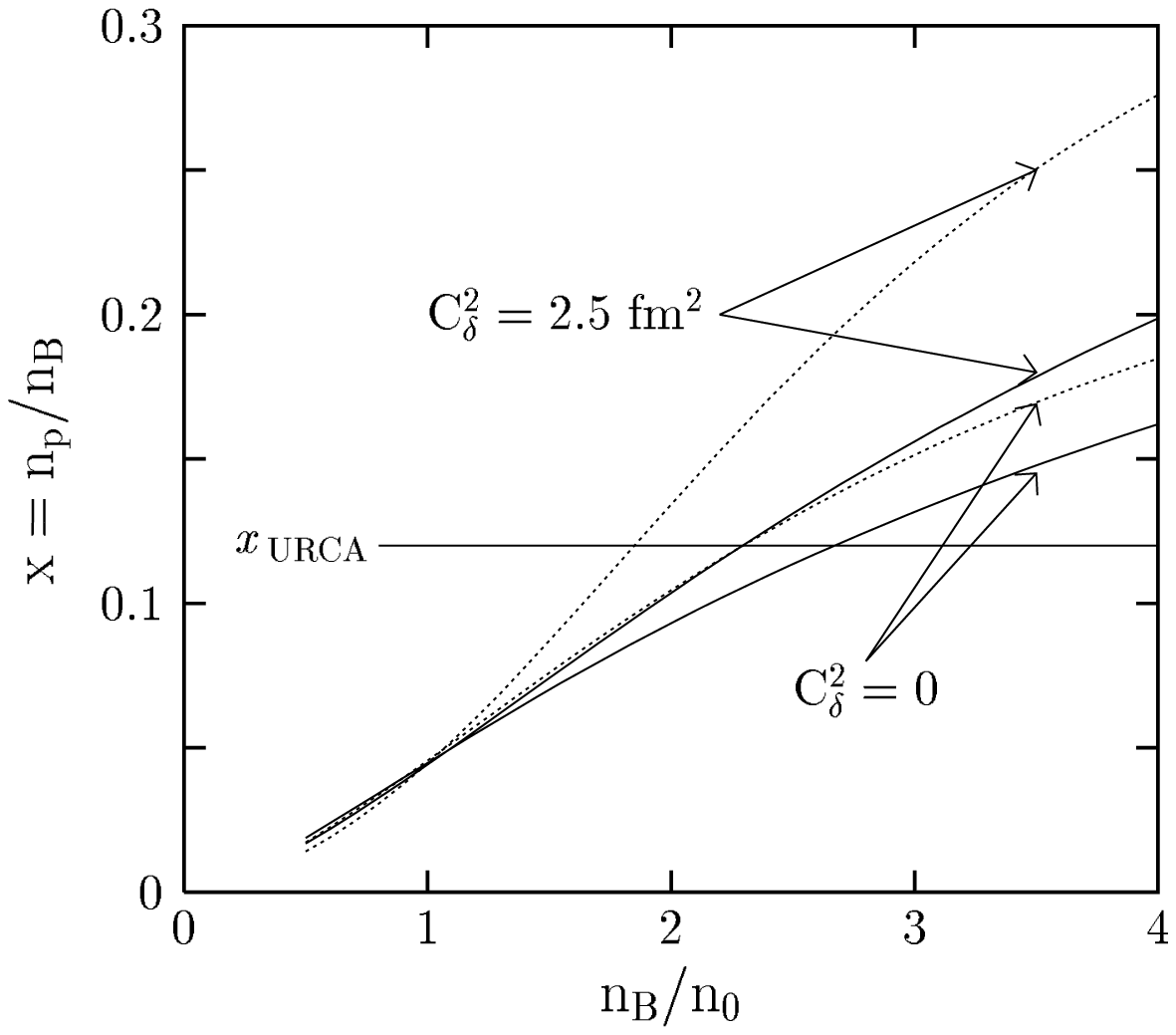}
\ni Fig.6

\ni Proton fraction of neutron star matter in the presence of
the $\delta$-field and with no $\delta$-field. Solid and dotted
curves correspond to the soft and stiff equation of state,
respectively. 

\vfill
\end